\documentclass[11pt,twoside]{article} 
\usepackage{acta-info}
\usepackage{euler}
\usepackage{graphics}
\usepackage{epstopdf}
\newcommand{\vol}{\textbf{2,} 2 (2010) 168--183}   %% to be completed by the editor, do not modify it
\newcommand{\amss}{\amssubj{%
97U50
}}     %% Source:   http://www.ams.org/msc/
\newcommand{\CRs}{\CRsubj{%
K.3.1
}}   %% Source: http://www.acm.org/about/class/1998
\newcommand{\keyw}{\keywords{%
computerized adaptive testing, item response theory
}}

\begin{document}

\title{% 
%%Write here the title of your paper, as
Computerized adaptive testing: implementation issues
}
\maketitle
%% SINGLE AUTHOR. If you are a single author, please, use the following command and delete 
%%                the \twoauthors command completely. 

%% TWO AUTHORS. If there are two authors, please, use the following command and delete 
%%              the \oneauthor command completely.  

\twoauthors{%
%% Write here the first author's name usinf \href command.
{\href{http://www.ms.sapientia.ro/~manyi/}{Margit Antal}}
}{%
 %% Write here the first author's affiliation including maybe his address. You can use \\ for line breaks.
{\href{http://www.emte.ro}{Sapientia Hungarian University of Transylvania}, 
\href{http://www.ms.sapientia.ro}{Department of \\ Mathematics and Informatics} \\ T\^{\i}rgu Mure\c s} 
}{%
 %% Write here the first author's email.
\href{mailto:manyi@ms.sapientia.ro}{manyi@ms.sapientia.ro} 
}{%
 %% Write here the second author's name.
{Levente Er\H{o}s} 
}{%
 %% Write here the second author's affiliation including maybe his address. You can use \\ for line breaks.
{\href{http://www.emte.ro}{Sapientia Hungarian University of Transylvania},  \href{http://www.ms.sapientia.ro}{Department of \\ Mathematics and Informatics}\\T\^{\i}rgu Mure\c s} 
}{%
 %% Write here the second author's email.
\href{ideges@gmail.com}{ideges@gmail.com} 
}

\oneauthor{%
 %% Write here the name of the author and this homepage address using \href command  
{Attila Imre} 
}{%
%% Write here your affiliation and its homepage address including maybe your address using \href command. 
%% You can use \\ for line breaks.
{\href{http://www.emte.ro}{Sapientia Hungarian University of Transylvania}, \href{http://www.ms.sapientia.ro}{ Department of Human Sciences}\\T\^{\i}rgu Mure\c s} 
}{%
 %% Write here your email.
 \href{imatex@ms.sapientia.ro}{imatex@ms.sapientia.ro}
}

%% Short name of the authors and short title, to be included in heading.
\short{%
%% Write here the short name of authors, using commas.
M. Antal, L. Er\H{o}s, A. Imre
}{%
%%Write here the short title
Computerized adaptive testing: implementation issues
}

\begin{abstract}
One of the fastest evolving field among teaching and learning research is students' performance evaluation. Computer based testing systems are increasingly adopted by universities. However, the implementation and maintenance of such a system and its underlying item bank is a challenge for an inexperienced tutor. Therefore, this paper discusses the advantages and disadvantages of Computer Adaptive Test (CAT) systems compared to Computer Based Test systems. Furthermore, a few item selection strategies are compared in order to overcome the item exposure drawback of such systems. The paper also presents our CAT system along its development steps. Besides, an item difficulty estimation technique is presented  based on data taken from our self-assessment system.
\end{abstract}

\newpage

\section{Introduction}

One of the fastest evolving field among teaching and learning research is students' performance evaluation. Web-based educational systems with integrated computer based testing are the easiest way of performance evaluation, so they are increasingly adopted by universities \cite{BARLA, BAYLARI, LILLEY}. With the rapid growth of computer communications technologies, online testing is becoming more and more common. Moreover, limitless opportunities of computers will cause the disappearance of Paper and Pencil (PP) tests. Computer administered tests present multiple advantages compared to PP tests. First of all, various multimedia can be attached to test items, which is almost impossible in PP tests. Secondly,  test evaluation is instantaneous. Moreover, computerized self-assessment systems can offer various hints, which  help students' exam preparation.

This paper is structured in more sections. Section 2 presents Item Response Theory (IRT) and discusses the advantages and disadvantages of adaptive test systems. Section 3 is dedicated to the implementation issues. The presentation of the item bank is followed by simulations for item selection strategies in order to overcome the item exposure drawback. Then the architecture of our web-based CAT system is presented, which is followed by a proposal for item difficulty estimation. Finally, we present further research directions and give our conclusions.

\section{Item Response Theory }

Computerized test systems reveal new testing opportunities. One of them is the adaptive item selection tailored to the examinee's ability level, which is estimated iteratively through the answered test items. Adaptive test administration consists in the following steps:  (i) start from an initial ability level, (ii) selection of the most appropriate test item  and (iii) based on the examinee's answer re-estimation of their ability level. The last two steps are repeated until some ending conditions are satisfied. Adaptive testing  research started in 1952 when Lord made an important observation: ability scores are test independent whereas observed scores are test dependent \cite{HAMBLETON}. The next  milestone was in 1960 when George Rasch described  a few item response models in his book \cite{RASCH}. One of the described models, the one-parameter logistic model, became known as the Rasch model. The next decades brought many new applications based on Item Response Theory. 

In the following we present the three-parameter logistic model. The basic component of this model is the item characteristic function:
\begin{equation}
P(\Theta )= c+\frac{(1-c)}{1+ e^{-Da(\Theta -b)}},
\label{eq3PL}
\end{equation} 
where $ \Theta $ stands for the examinee's ability, whose theoretical range is from $ - \infty $  to  $ \infty $, but practically the range $-3$ to $+3$ is used. The three parameters are: $a$, discrimination; $b$, difficulty; $c$,  guessing.  Discrimination determines how well an item differentiates students near an ability level. Difficulty shows the place of an item along the ability scale, and guessing represents the probability of guessing the correct answer of the item \cite{BAKER}. Therefore guessing for a true/false item is always 0.5. $P(\Theta)$  is the probability of a correct response to the item  as a function of ability level \cite{HAMBLETON}. $D$  is a scaling factor and typically the value $1.7$ is used. 
  
\begin{figure}[t]
\begin{center}\includegraphics[scale=0.7]{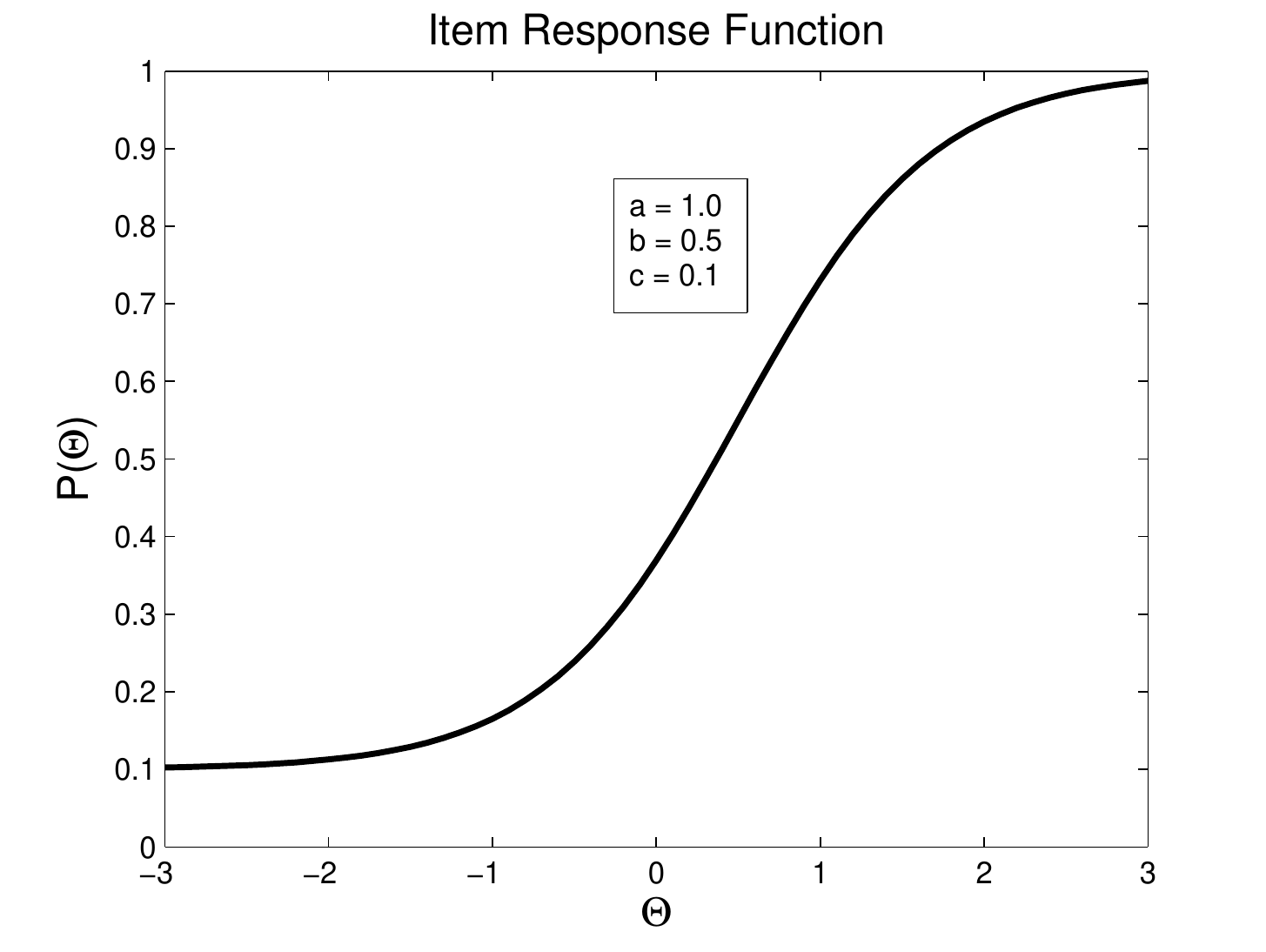}\end{center}
\caption{\label{IRF}A three-parameter logistic model item characteristic function} 
\end{figure}

\begin{figure}[t]
\begin{center}\includegraphics[scale=0.7]{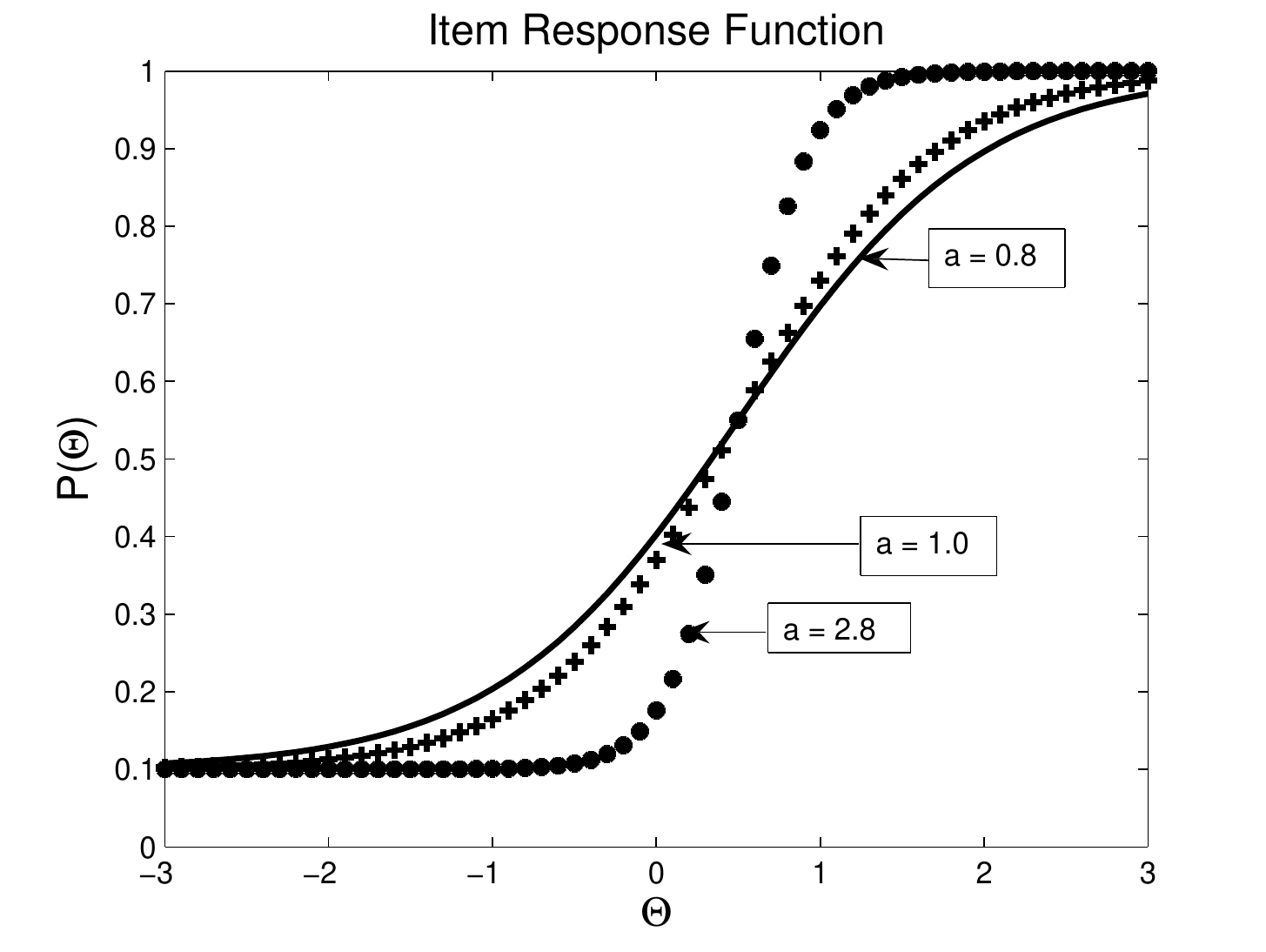}\end{center}
\caption{\label{ability3}Item characteristic functions} 
\end{figure}

Figure \ref{IRF} shows item response function for an item having parameters $a=1, \quad b=0.5, \quad c=0.1$.
For a deeper understanding of the discrimination parameter, see Figure \ref{ability3}, which illustrates three different items with the same difficulty ($b=0.5$) and guessing ($c=0.1$) but different discrimination parameters. The steepest curve corresponds to the highest discrimination ($a=2.8$), and in the middle of the curve the probability of correct answer changes very rapidly as ability increases \cite{BAKER}. 

The one- and two-parameter logistic models can be obtained from equation (\ref{eq3PL}), for example setting $c=0$  results in the two-parameter model, while setting $c=0$ and $a=1$ gives us the one-parameter model.

Compared to the classical test theory, it is easy to realize the benefits of the former, which is able to propose the most appropriate item, based on item statistics reported on the same scale as ability \cite{HAMBLETON}. 

Another component of the IRT model is the item information function, which shows the contribution of a particular item to the assessment of ability \cite{HAMBLETON}. Item information functions are usually bell shaped functions, and in this paper we used  the following (recommended in \cite{RUDNER}): 
\begin{equation}
I_{i}(\Theta)=\frac{P_{i}^{'}(\Theta)^2}{P_{i}(\Theta)(1-P_{i}(\Theta))},
\label{eqI}
\end{equation}
where $P_{i}(\Theta)$ is the probability of a correct response to item $i$ computed by equation (\ref{eq3PL}), $P_{i}^{'}(\Theta)$ is the first derivative of $P_{i}(\Theta)$, and $I_{i}(\Theta)$ is the item information function for item $i$.

\begin{figure}[t]
\begin{center}\includegraphics[scale=0.7]{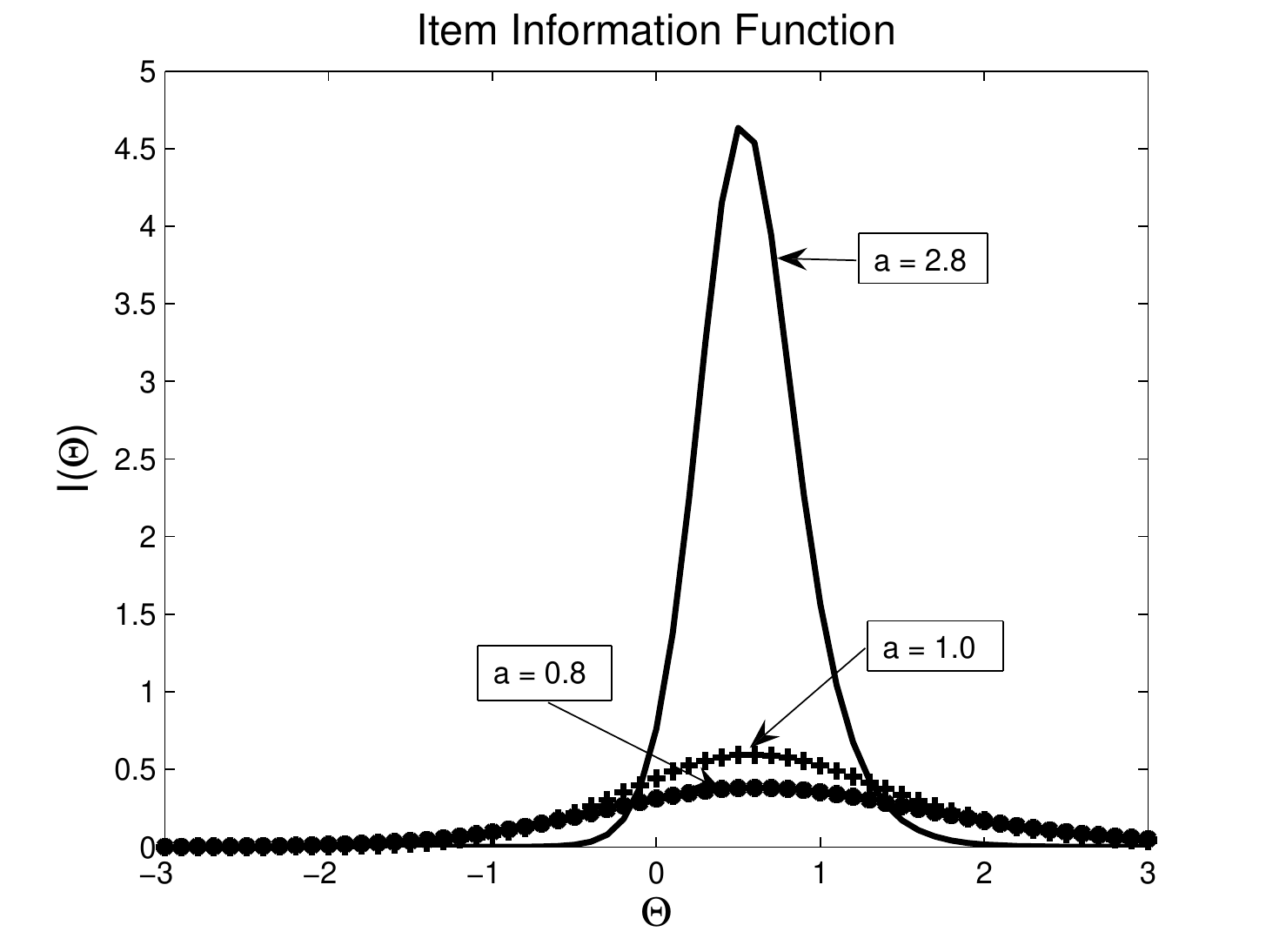}\end{center}
\caption{\label{iteminfo3}Item information functions} 
\end{figure}

High discriminating power items are the best choice  as shown in  Figure \ref{iteminfo3}, which illustrates the item information functions for the three items shown in Figure \ref{ability3}. All three functions are centered around the ability $\Theta=0.5$, which is the same as the item difficulty.

\begin{figure}
\begin{center}\includegraphics[scale=0.6]{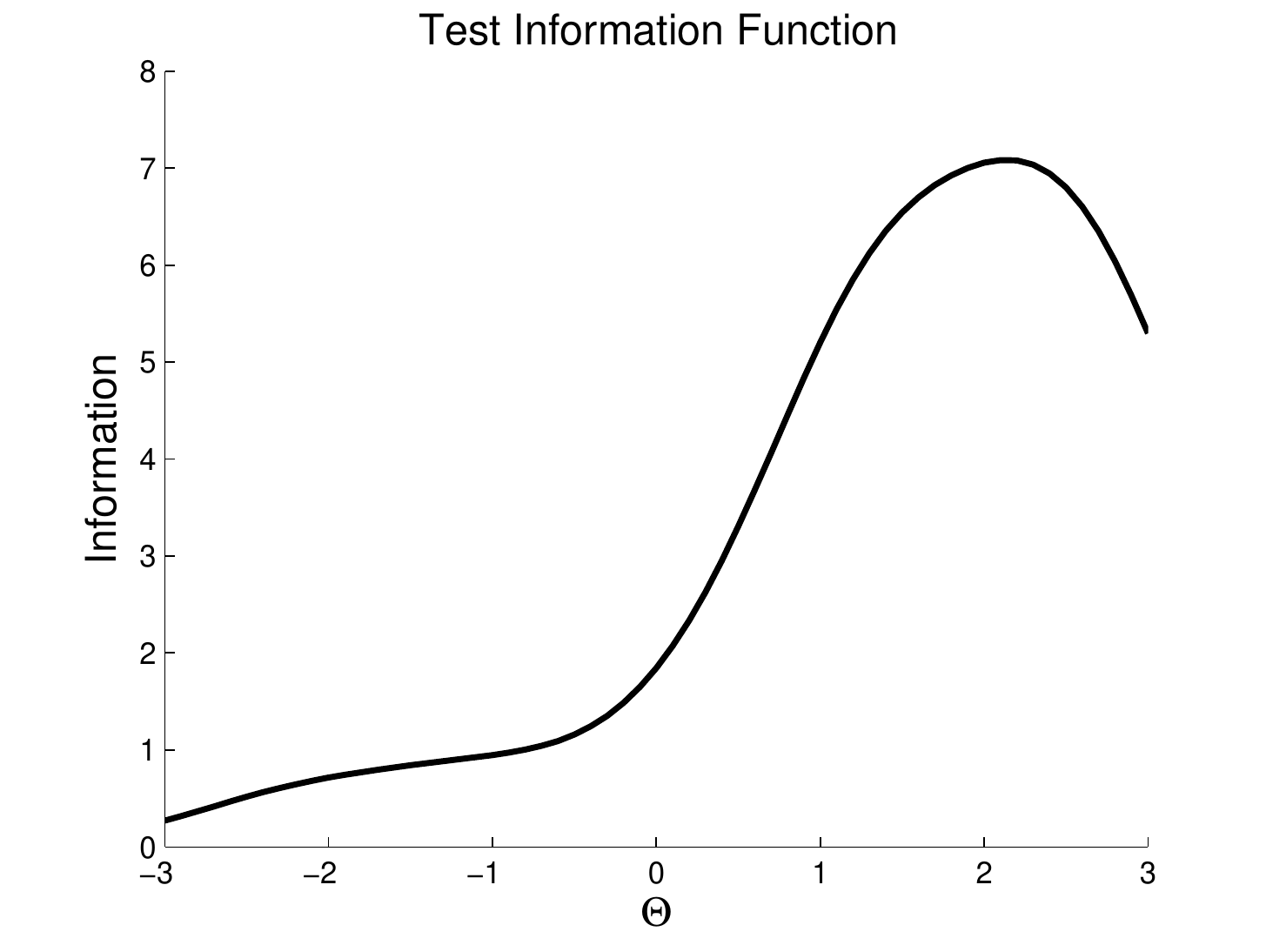} \end{center}          
\caption{\label{fig:smart}Test information function for a 20-item test generated for  high ability students} 
\end{figure}

\begin{figure}
\begin{center}    \includegraphics[scale=0.6]{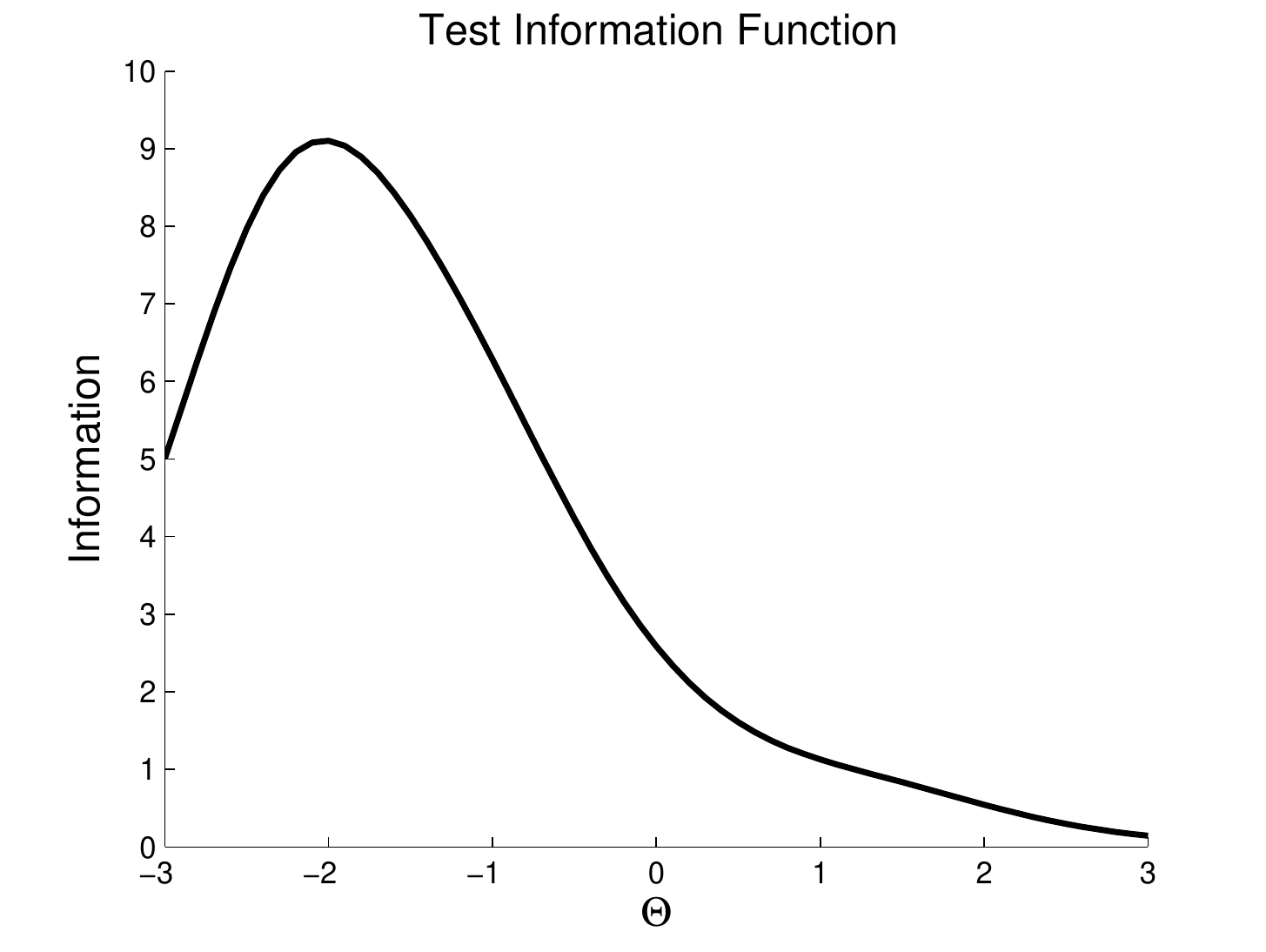} \end{center}
\caption{\label{fig:dull} Test information function for a 20-item test generated for low ability students} 
\end{figure}

Test information function $I_{rr}$ is defined as the sum of item information functions. Two such functions are shown for a 20-item test selected by our adaptive test system: one for a high ability student (Figure \ref{fig:smart})  and another for a low ability student (Figure \ref{fig:dull}). The test shown in Figure \ref{fig:smart} estimates students' ability near $\Theta=2.0$, while the test in Figure \ref{fig:dull} at $\Theta= -2.0$.

Test information function is also used for ability estimation error computation as shown in the following equation:

\begin{equation}
SE(\Theta) = \frac{1}{\sqrt{I_{rr}}}.
\end{equation}
This error is associated with maximum likelihood ability estimation and is usually used for the stopping condition of adaptive testing.

\begin{figure}
\begin{center}    \includegraphics[scale=0.6]{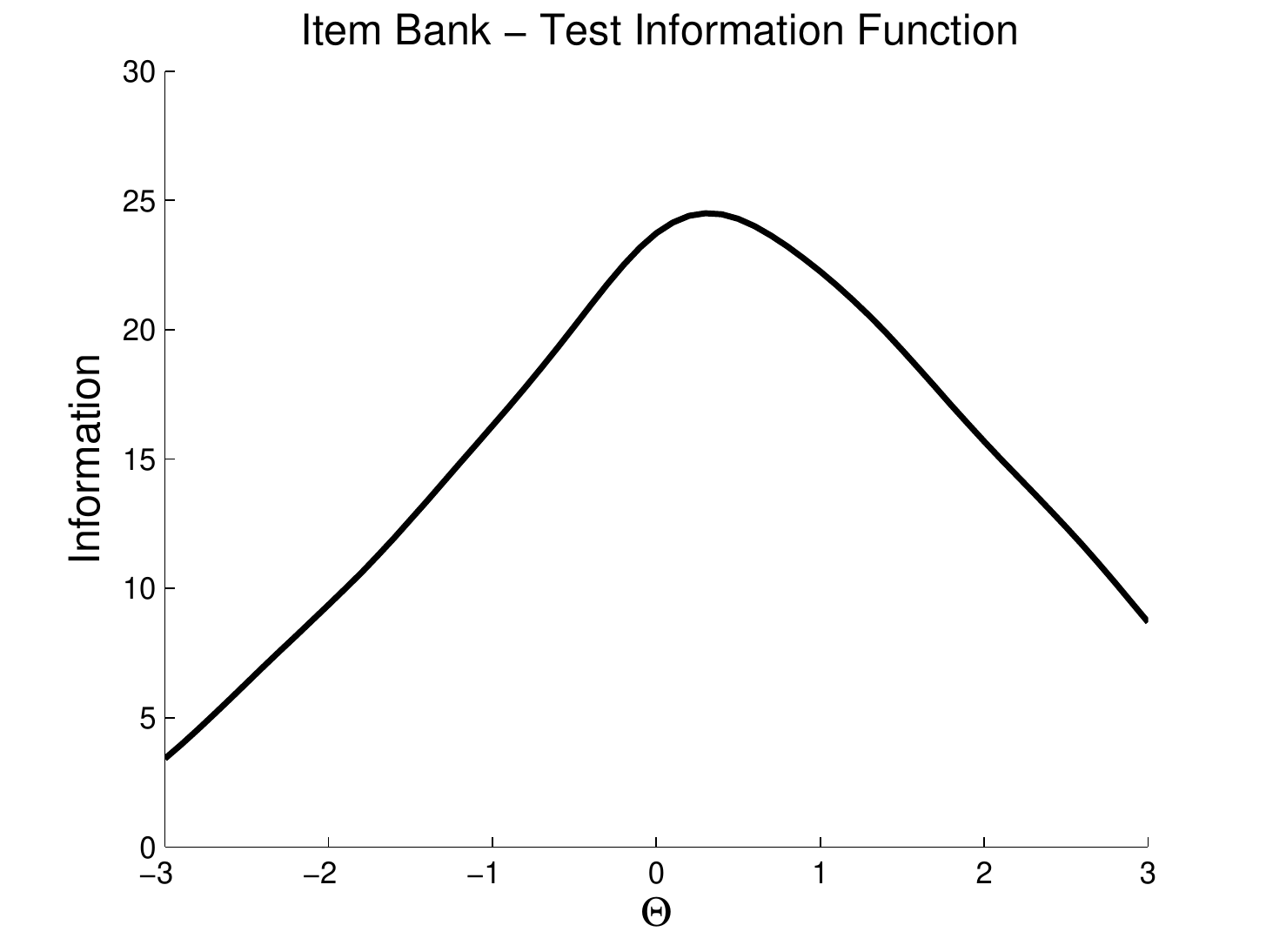} \end{center}
\caption{\label{fig:all} Test information function for all test items} 
\end{figure}

For learner proficiency estimation Lord proposes an iterative approach \cite{LORD}, which is a modified version of the Newton-Raphson iterative method for solving equations. This approach starts with an initial ability estimate (usually a random value). After each item the ability is adjusted based on the response given by the examinee. For example, after $n$ questions the estimation is made according to the following equation:

\begin{equation}
\Theta_{n+1} = \Theta_{n}+\frac{\displaystyle\sum_{i=1}^{n}S_i(\Theta_n)}{\displaystyle\sum_{i=1}^{n}I_i(\Theta_n)}
\label{eqTheta}
\end{equation}
where $S_i(\Theta)$ is computed using the following equation:

\begin{equation}
S_i(\Theta)=(u_i-P_i(\Theta))\frac{P_i^{'}(\Theta)}{P_i(\Theta)(1-P_i(\Theta))}.
\label{eqS}
\end{equation}
In equation (\ref{eqS}) $u_i$ represents the correctness of the $i$th  answer, which is $0$ for incorrect and $1$ for correct answer. $P_{i}(\Theta)$ is the probability of correct answer for the $i$th item having the ability $\Theta$ (equation (\ref{eqI})), and  $P_{i}^{'}(\Theta)$ is its  first derivative.

\subsection{Advantages}

In adaptive testing the best test item is selected at each step: the item having maximum information at the current estimate of the examinee's proficiency. The most important advantage of this method is that high ability level test-takers are not bored with easy test items, while low ability ones are not faced with difficult test items. A consequence of adapting the test to the examinee's ability level is  that the same measurement precision can be realized with fewer test items.

\subsection{Disadvantages}

Along with the advantages offered by IRT, there are some drawbacks as well. The first drawback is the impossibility to estimate the ability in case of all correct or zero correct responses. These are the cases of either very high or very low ability students. In such cases the test item administration must be stopped after administering a minimum number of questions.

The second drawback is that the basic IRT algorithm is not aware of the test content, the question selection strategy does not take into consideration to which topic a question belongs. However, sometimes this may be a requirement for generating tests assessing certain topics in a given curriculum. Huang proposed a content-balanced adaptive testing algorithm \cite{HUANG}. Another solution to the content balancing problem is the testlet approach proposed by Wainer and Kiely \cite{WAINER}.  A testlet is a group of items from a single curriculum topic, which is developed as a unit. If an adaptive algorithm selects a testlet, then all the items belonging to that testlet will be presented to the examinee.

The third drawback, which is also the major one, is that IRT algorithms require serious item calibration. Despite the fact that the first calibration method was proposed by Alan Birnbaum in 1968 and has been implemented in computer programs such as BICAL (Wright and Mead, 1976) and LOGIST (Wingersky, Barton and Lord, 1982), the technique needs real measurement data in order to accurately estimate the parameters of the items. However, real measurement data are not always available for small educational institutions.

The fourth drawback is that several items from the item bank will be overexposed, while other test items will not be used at all. This requires item exposure control strategies.  A good review of these strategies can be found in \cite{GEORGIADOU}, discussing the strengths and weaknesses of each strategy. Stocking \cite{STOCKING} made one of the first overviews of item exposure control strategies and classified them in two groups: (i) methods using a random component along the item selection method and (ii) methods using a parameter for each item to control its exposure.  Randomization strategies control the frequency of item administration by selecting the next item from a group of items (e.g. out of the 5 best items). The second item exposure control strategy uses an exposure control parameter. In case of an item selection---due to its maximum information for the examinee's ability level---, the item will be administered only if its exposure control parameter  allows it.

\section{CAT implementation}

\subsection{The item bank}

We have used our own item bank from our traditional computer based test system "Intelligent" \cite{ANTAL}. The item bank  parameters ($a$ - discrimination, $b$ - difficulty, $c$ - pseudo guessing) were initialized by the tutor. We used 5 levels of difficulty from very easy to very difficult, which were scaled to the [-3,3] interval. The guessing parameter of an item was initialized by the ratio of the number of possible correct answers to the total number of possible answers. For example, it is 0.1 for an item having two correct answers out of five possible answers. Discrimination is difficult to set even for a tutor, therefore we used $a=1$ for each item. 

\subsection{Simulations}

\begin{figure}
\begin{center}    \includegraphics[scale=0.6]{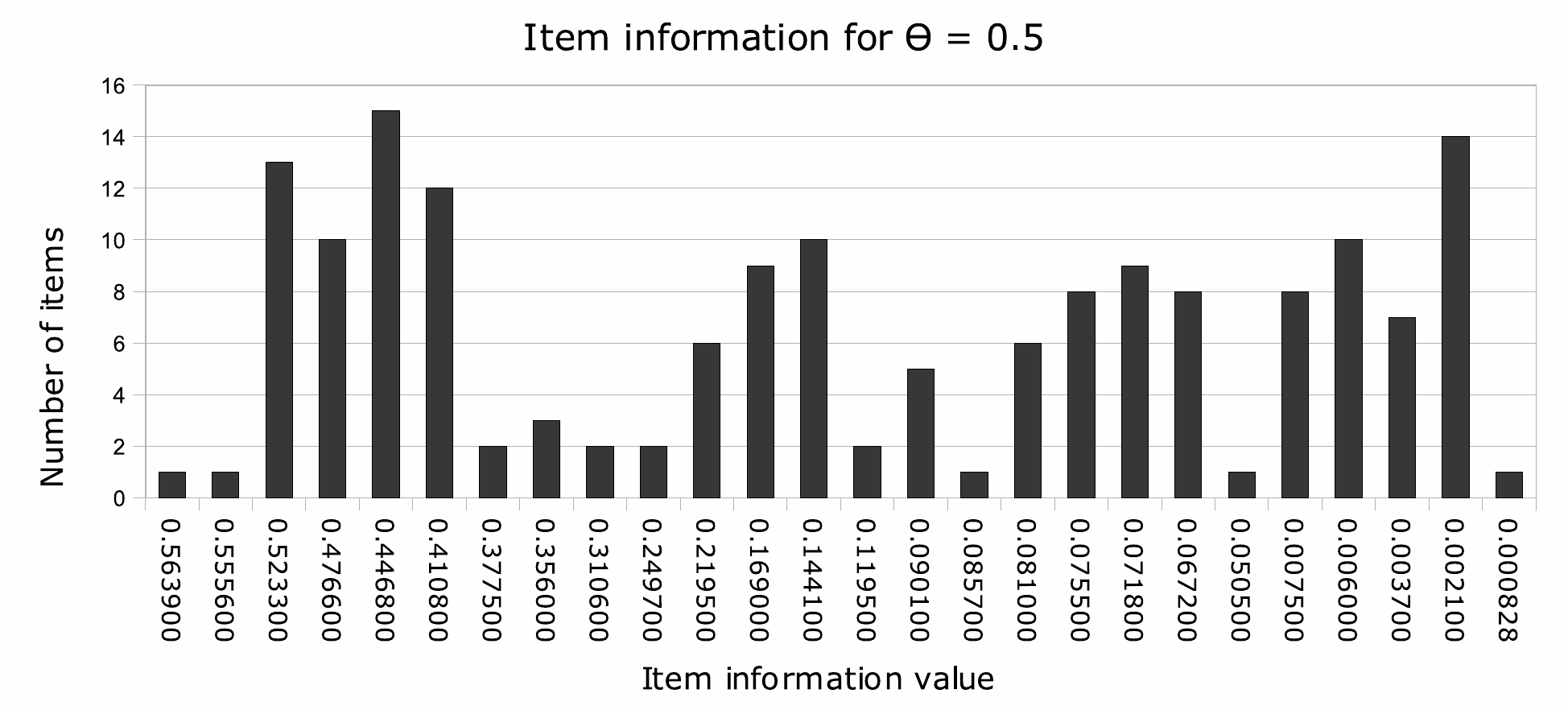} \end{center}
\caption{\label{fig:II_clusters} Item information clusters and their size} 
\end{figure}

In our implementation we have tried to overcome the disadvantages of IRT. We started to administer items adaptively only after the first five items. Ability ($\Theta$) was initialized based on the number of correct answers given to these five items, which  had been selected to include all levels of difficulty.

We used randomization strategies to overcome item exposure. Two randomization strategies were simulated. In the first one we selected the top ten items, i.e. the ten items having the highest item information. However, this is better than choosing the single best item, thus one must pay attention to the selection of the top ten items. 
There may be more items having the same item information for a given ability, therefore it is not always the best strategy choosing the first best item from a set of items with the same item information. To overcome this problem, in the second randomization strategy we computed the item information for all items that were not presented to the examinee and clustered the items having the same item information. The top ten items were selected using the items from these clusters. If the best cluster had less than ten items, the remainder items were selected from the next best cluster. If the best cluster had more than ten items, the ten items were selected randomly from the best cluster. 
For example, Figure \ref{fig:II_clusters} shows the 26 clusters of item information values constructed from 171 items for the ability of $\Theta=0.5$. The best 10 items were selected by taking the items from the first two clusters (each having exactly 1 item) and selecting randomly another 8 items out of 13 from the third cluster.

\begin{figure}[t]
\begin{center}    \includegraphics[scale=0.9]{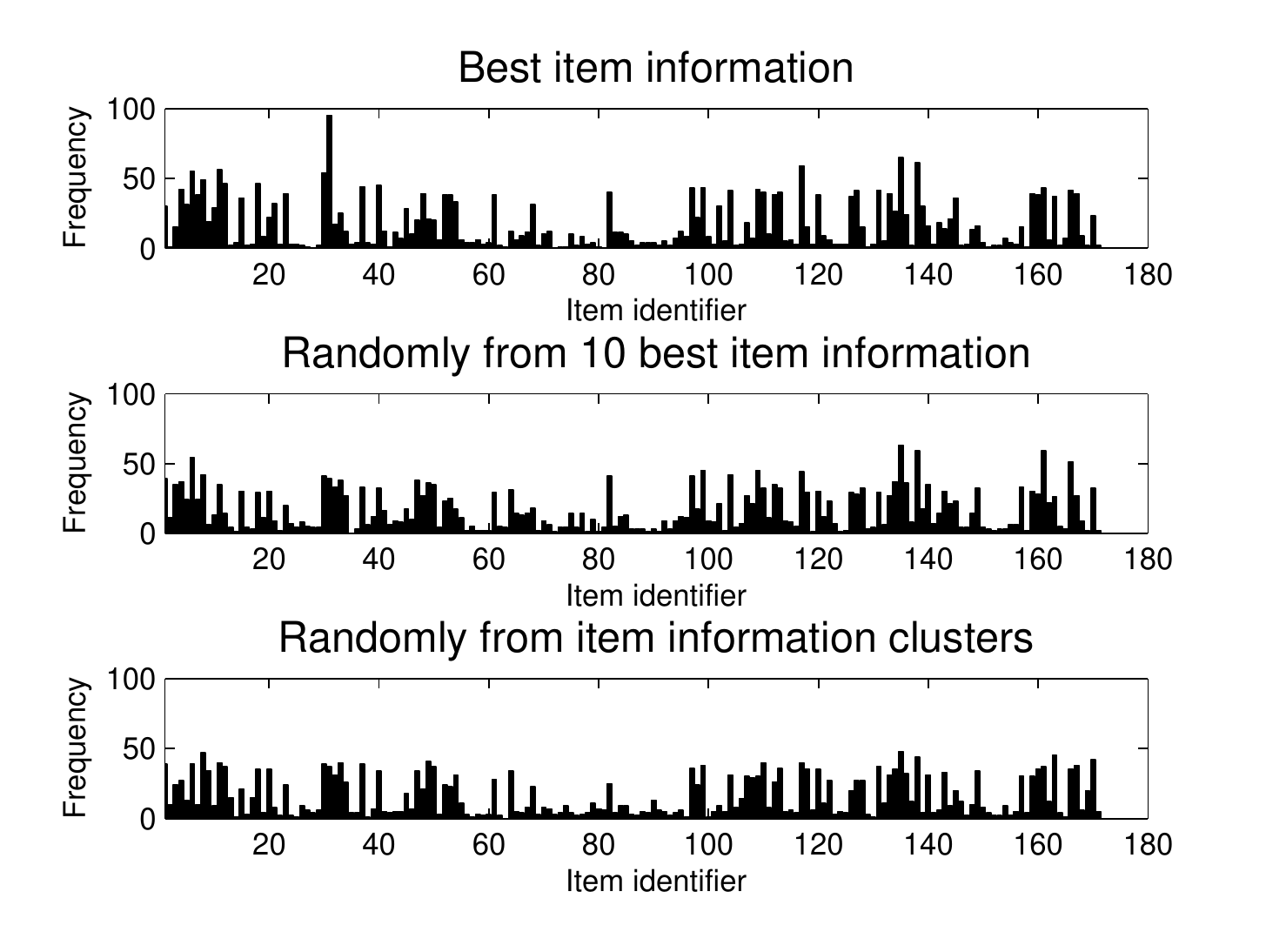} \end{center}
\caption{\label{fig:plotItemExposure} Item exposure with or without randomization control strategies} 
\end{figure}

Figure \ref{fig:plotItemExposure} shows the results from a simulation where we used an item bank with 171 items (test information function is shown in Figure \ref{fig:all} for all the 171 items), and we simulated 100 examinees using three item selection strategies: (i) best item (ii) random selection from the 10 best items (iii) random selection from the 10 best items and clustering. The three series in figure \ref{fig:plotItemExposure} are the frequencies of items obtained from the 100 simulated adaptive tests. Tests were terminated either when the number of administered items had exceeded 30 or the ability estimate had  fallen outside the ability range. The latter were necessary for very high and very low ability students, where adaptive selection could not be used \cite{BAKER}. The examinee's answers were simulated by using a uniform random number generator, where the probability of correct answer was set to be equal to the probability of incorrect answer.

In order to be able to compare these item exposure control strategies, we computed the standard deviance of the frequency series shown in Figure \ref{fig:plotItemExposure}. The standard deviance is $\sigma = 17.68$ for the first series not using any item exposure control, it is $\sigma = 14.77$ for the second one, whereas for the third one is $\sigma = 14.13$. It is obvious that the third series is the best from the viewpoint of item exposure. Consequently, we will adopt this strategy in our distributed CAT implementation.

\subsection{Distributed CAT}
\begin{figure}[t]
\begin{center}    \includegraphics[scale=0.8]{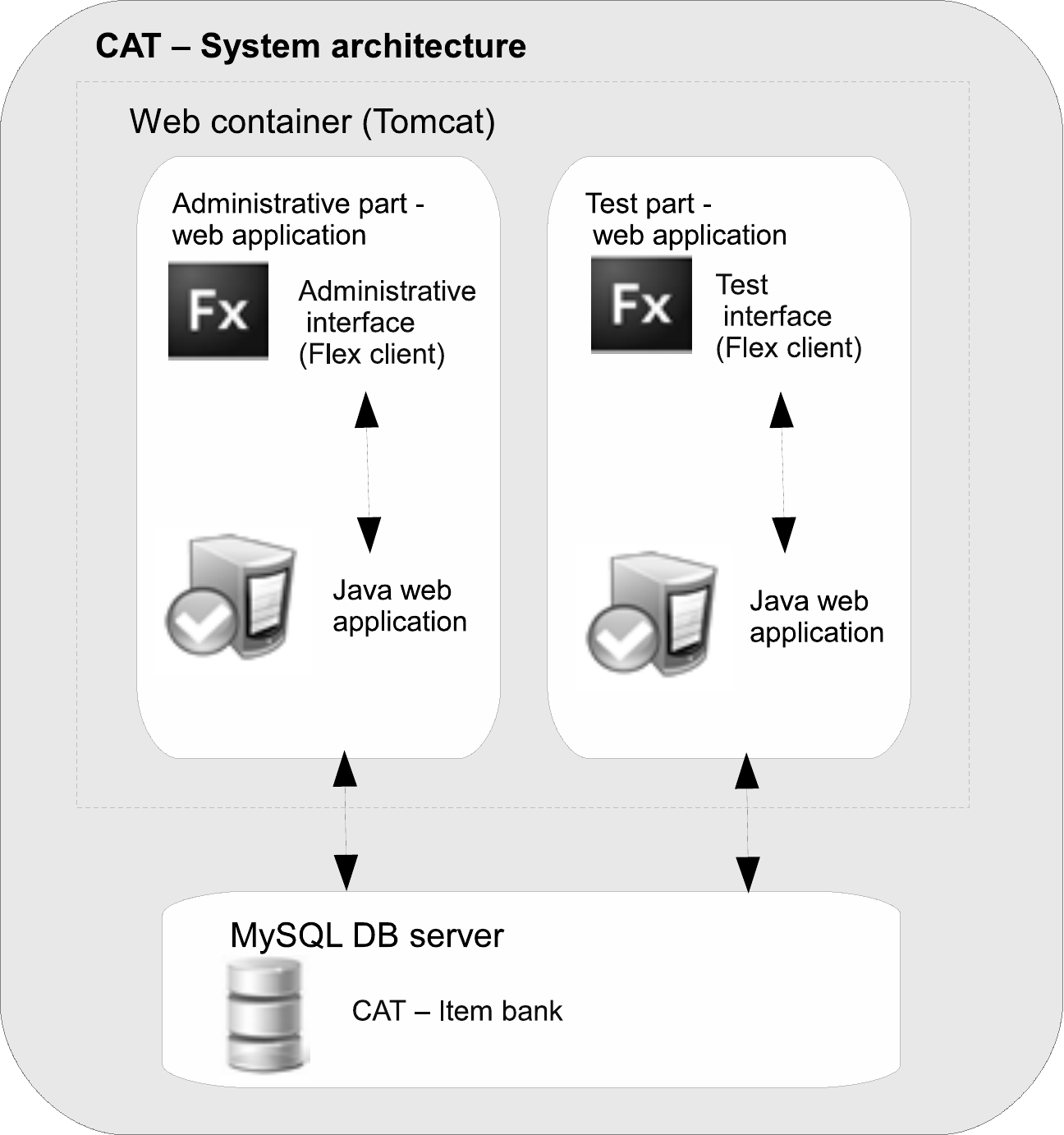} \end{center}
\caption{\label{fig:cat_architecture} CAT-architecture} 
\end{figure}

After the Matlab simulations we implemented  our CAT system as a distributed application, using Java technologies on the server side and Adobe Flex on the client side. The general architecture of our system is shown in Figure \ref{fig:cat_architecture}. The administrative part is responsible for item bank maintenance, test scheduling, test results statistics and test termination criteria settings. In the near future we are planning to add an item calibration module. 

The test part is used by examinees, where questions are administered according to settings. After having finished the test, the examinee may view both their test results and knowledge report.

\subsection{Item difficulty estimation}

Due to the lack of measurement data necessary for item calibration, we were not able to calibrate our item bank. However, 165 out of 171 items of our item bank were used in our self-assessment test system "Intelligent" in the previous term. Based on the data collected from this system, we propose a method for difficulty parameter estimation. Although there were no restrictions in using the self-assessment system, i.e. users could have answered an item several times, we consider that the first answer of each user could be  relevant to the difficulty of the item. 

\begin{figure}[t]
\begin{center}    \includegraphics[scale=0.8]{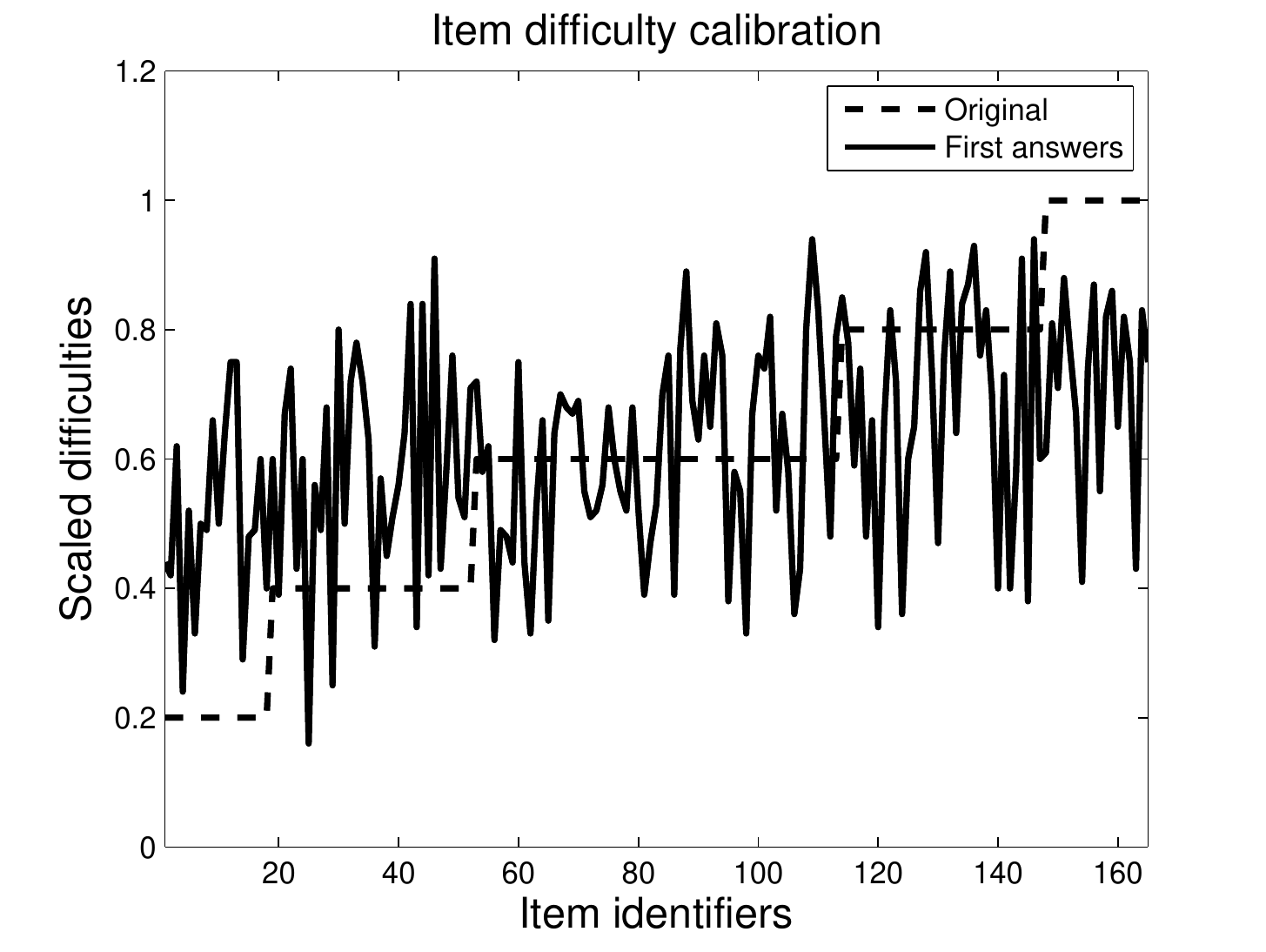} \end{center}
\caption{\label{fig:itemcalibration1} Item difficulty calibration} 
\end{figure}

Figure \ref{fig:itemcalibration1} shows the original item difficulty (set by the tutor) and the difficulty estimated by the first answer of each user. The original data series uses 5 difficulty levels scaled to the $[0, 1]$ interval. The elements of the ``first answers'' series were computed by the equation:  $\frac{all\_incorrect\_answers}{all\_answers}$. We computed the mean difficulty for both series, and we obtained 0.60 for the original one and 0.62 for the estimated one. Conspicuous differences were found at the \textit{very easy} and \textit{very difficult} item difficulties.

\section{Further research}

At present we are working on the parameter estimation part of our CAT system. Although there are several item parameter calibration programs, this task must be taken very seriously because it influences measurement precision directly. Item parameter estimation error is an active research topic, especially for fixed computer tests. For adaptive testing, this problem has been addressed by paper \cite{LINDENGLAS}. 

Researchers have empirically observed that examinees suitable for item difficulty estimations are almost useless when estimating item discrimination. Stocking \cite{STOCKING1990} analytically derived the relationship between the examinee's ability and the accuracy of maximum likelihood item parameter estimation. She concluded that high ability examinees contribute more to difficulty estimation of difficult and very difficult items and less on easy and very easy items. She also concluded that only low ability examinees contribute to the estimation of guessing parameter and examinees, who are informative regarding item difficulty estimation, are not good for item discrimination estimation. Consequently, her results seem to be useful in our item calibration module.

\section{Conclusions}

In this paper we have described a computer adaptive test system based on Item Response Theory along its implementation issues. Our CAT system was implemented after one year experience with a computer based self-assessment system, which proved useful in configuring the parameters of the items. We started with the presentation of the exact formulas used by our working CAT system, followed by some simulations for item selection strategies in order to control item overexposure. We also presented the exact way of item parameter configuration based on the data taken from the self-assessment system.

Although we do not present measurements on a working CAT system, the implementation details presented in this paper could be useful for small institutions planning to introduce such a system for educational measurements on a small scale.

In the near future we would like to add an item calibration module to the administrative part of the system, taking into account the limited possibilities of small educational institutes.

\bigskip
\rightline{\emph{Received: August 23, 2010 {\tiny \raisebox{2pt}{$\bullet$\!}} Revised: November 3, 2010}} %% to be completed by the editor

\end{document}